\documentclass[preprint,twocolumn,3p]{elsarticle}
\usepackage{numcompress}
\usepackage{times}
\usepackage[hyphens]{url}  
\usepackage{latexsym}
\usepackage{amsmath}
\usepackage{hyperref}
\usepackage{xcolor}
\usepackage{booktabs}  

\biboptions{authoryear}

\begin{document}
\begin{frontmatter}

\title{How sustainable is ``common" data science in terms of power consumption?}

\author[1]{Bjorge Meulemeester}
\ead{bjorge.meulemeester@uantwerpen.be}
\author[1]{prof. D. Martens}
\ead{david.martens@uantwerpen.be}
\address[1]{Prinsstraat 13, 2000 Antwerpen, BEL}

\begin{abstract}
Continuous developments in data science have brought forth an exponential increase in complexity of machine learning models. Additionally, data scientists have become ubiquitous in the private market, academic environments and even as a hobby. All of these trends are on a steady rise, and are associated with an increase in power consumption and associated carbon footprint. The increasing carbon footprint of large-scale advanced data science has already received attention, but the latter trend has not. This work aims to estimate the contribution of the increasingly popular “common” data science to the global carbon footprint. To this end, the power consumption of several typical tasks in the aforementioned common data science tasks will be measured and compared to: large-scale “advanced” data science, common computer-related tasks, and everyday non-computer related tasks. This is done by converting the measurements to the equivalent unit of “km driven by car”. Our main findings are: “common” data science consumes $2.57$ more power than regular computer usage, but less than some common everyday power-consuming tasks such as lighting or heating; large-scale data science consumes substantially more power than common data science.
\end{abstract}


\end{frontmatter}

\begin{abstract}
Continuous developments in data science have brought forth an exponential increase in complexity of machine learning models. Additionally, data scientists have become ubiquitous in the private market, academic environments and even as a hobby. All of these trends are on a steady rise, and are associated with an increase in power consumption and associated carbon footprint. The increasing carbon footprint of large-scale advanced data science has already received attention, but the latter trend has not. This work aims to estimate the contribution of the increasingly popular “common” data science to the global carbon footprint. To this end, the power consumption of several typical tasks in the aforementioned common data science tasks will be measured and compared to: large-scale “advanced” data science, common computer-related tasks, and everyday non-computer related tasks. This is done by converting the measurements to the equivalent unit of “km driven by car”. Our main findings are: “common” data science consumes $2.57$ more power than regular computer usage, but less than some common everyday power-consuming tasks such as lighting or heating; large-scale data science consumes substantially more power than common data science.
\end{abstract}

\section{Introduction}
Data science and Artificial Intelligence (AI) have become a fundamental aspect of our digital world; be it for marketing strategies, uncovering shopping patterns of clients, computer-mediated language comprehension, detecting cardiac arrhythmias or simply predicting the weather. Deep neural networks have experienced a steep increase in complexity, and correspondingly a steep increase in the hardware required to train these models \citep{anthony2020carbontracker}. As the hardware requirements have grown exponentially, so has the power consumption attributed to these hardware requirements. Large models are commonly trained in big datacenters, powered by various types of energy \citep{CompanyRenewableGreenpeace}. It is estimated that these datacenters currently account for about $1\%$ of the worldwide electricity use \citep{masanet2020}. Their projected energy consumption and carbon emission for the upcoming years is subject of debate, and estimates range between the status quo \citep{masanet2020} up to $8\%$ of global carbon emission by 2030 \citep{cao2022systematic}. Major takeaways from this discussion is that datacenters are gradually shifting towards green energy \citep{ICTGlobalEmission}, are getting increasingly efficient \citep{masanet2020}, but are simultaneously being used more and more for hyperscale data science \citep{IEA} and, despite the shift towards green energy in datacenter’s infrastructure being widely adopted \citep{bashroush2020beyond}, measures targeting other areas, such as inefficiencies in servers, circular economy, and heat reuse, are still lagging \citep{RTEIL2022100618}.

These values represents the steep increase in data center demand and their usages, but does not take into account the increase in ``common" data science jobs that do not require these large data centers. The term common data science will be used in this work to denote any data science that can be performed by one (or a team of) data scientist(s) on their home computer/laptop.

Data science entails more than huge complex neural networks and big datacenters; common data scientists are omnipresent in the private market in various fields. Data science related jobs often end up in the top 5 fastest growing jobs in terms of demand: according to LinkedIn, \textit{Machine learning engineer} is the $4^{th}$ fastest rising job in the U.S. \citep{fast_US_linkedin}, and the $2^{nd}$ fastest rising job in The Netherlands, Italy and the U.K. \citep{fast_EU_linkedin}. Other data science related jobs such as \textit{Data scientist} and \textit{Data engineer} tend to end up high on these lists as well. Does this increase in common ``everyday" data science also imply an increase in power consumption and carbon emission? \cite{masanet2020} estimate that in $2020$, powering digital devices (excluding smartphones) accounted for $~20\%$ of all ICT-related greenhouse gas emissions (GHGe), and communication networks account for $~25\%$; the sum of which is comparable to the estimate of data center GHGe of $~45\%$. The contribution to the global GHGe of household ICT devices, such as those used by common data scientists, should not be underestimated.

This work aims to compare the previously mentioned advanced data science, or ``big datacenter data science" to the latter common data science. The main question is if the latter common data science has a substantial contribution to the global GHGe. To this end, the carbon emissions of common data science will be estimated and compared to common computer tasks, common non-computer related everyday tasks and the aforementioned ``big datacenter" advanced data science. This is done by means of performing various tasks on a laptop while measuring their power consumption, comparing these values to one another, and to values found in literature.
\section{Methods}
In order to measure the carbon emission of different kinds of computer usage, three steps need to be taken:
\begin{enumerate}
    \item Measuring the power consumption and/or energy consumption of a single task.
    \item Converting the energy usage to carbon emission.
    \item Converting the carbon emission to an interpretable quantity, e.g. km driven by car.
\end{enumerate}

The ``everyday tasks, common data science tasks and advanced data science tasks that are to be compared will be explained in further detail in Section \ref{sec:measurements}.

\subsection{Assumptions}
\label{sec:assumptions}
Inevitably, assumptions need to be made in order to compare the measurements in a comprehensive manner. The measurements in this work will only be as truthful as the truth value of these assumptions. These will also be revisited in Section \ref{sec:assumptions_revisited}.
\begin{enumerate}
\item It suffices to measure the power consumption of the CPU to gain insight in the power consumption of performing various computer tasks.
\item The power demand of the hardware used in this work is representative for other hardware in common data science jobs, or can be extrapolated to other hardware types by considering the Thermal Design Power (TDP) of the CPU.
\item The authors' work efficiency and methods are representative for the work of other data scientists.
\item Measuring the power consumption of the hardware every 10 seconds suffices to give an accurate representation of the real-time power consumption of the hardware.
\item A car emits as much $gCO^2eq$ as the EU average from 2019.
\item Energy production emits as much $CO^2eq$ as the EU average from 2020.
\end{enumerate}

\subsection{Measuring the power consumption}
\label{measuring}
Measuring the power consumption and/or energy consumption of a task performed on the computer will be done by using Ubuntu's built-in powerstat command to sample the intel RAPL interface. This can be done by running the shell command
\begin{flalign}
    & \texttt{sudo powerstat -DRgf -d=<s> $\Delta_t$} \nonumber \\
    & \texttt{N > output.txt}
\end{flalign}
where \texttt{-D} enables the \texttt{-R} option, \texttt{-R} denotes sampling should be done on the RAPL interface, \texttt{-f} enables showing the average CPU frequency, \texttt{-d} denotes a delay before starting the measurement (in seconds), $\Delta_t$ denotes the time interval between samples (in seconds), \texttt{N} denotes the amount of samples and \texttt{-g} enables measuring the GPU power consumption as well. GPU power consumption and GPU programming will not be explored in this work, but was still measured for completeness.

Where possible, CarbonTracker \citep{anthony2020carbontracker} will also be used to measure energy consumption. CarbonTracker is designed to track and predict the energy consumption of training AI's, but can be used for any piece of code. It must be noted that CarbonTracker was designed for measuring the power consumption of training an AI during one or more epochs, and not shorter code snippets prevalent in common data science. Due to this, its accuracy may deviate, especially if the runtime of the code is short. For this reason, interpreting the data will be based on the RAPL measurements, and the CarbonTracker measurements will only be shown for completeness, and as a benchmark to check if it produces similar results. Deviating results will also be discussed in Section \ref{sec:discussion}.

\subsection{Converting}
\label{sec:converting}
In order to link the energy consumption to carbon emission, the EU-28 average greenhouse gas emission associated with electricity generation during 2020 \citep{EU28average} will be used, yielding an average emission per energy unit of:
\begin{equation}
    CO_2eq / energy = 230.7\ g / kWh
\end{equation}where $CO_2eq$ denotes the amounts of gram $CO_2$ that would yield a greenhouse effect of equal magnitude as the considered greenhouse gases, i.e. the seven greenhouse gases considered by the Kyoto Protocol \citep{protocol1997kyoto}: carbon dioxide ($CO_2$), methane ($NH_4$), nitrous oxide ($N_2O$),     hydrofluorocarbons ($HFCs$), perfluorocarbons ($PFCs$), sulphur hexafluoride ($SF_6$) and nitrogen trifluoride ($NF_3$).

In order to increase the interpretability of the $g CO_2eq$ quantity, they will be converted to the quantity ``km driven by car, as suggested by \cite{anthony2020carbontracker}. To do so, we will use the average GHGe of every registered car in the EU up until 2019 \citep{carco2}, yielding a distance per equivalent carbon emission of:
\begin{equation}
     8.17661488144\ m/gCO_2eq
\end{equation}

\subsection{Hardware \& software in this work}
All measurements were taken on a Dell Inspiron 15 7570 laptop with:
\begin{itemize}
    \item 7.5GB of RAM
    \item 64-bit Intel® Core™ i7-8550U CPU \\@ 1.80GHz × 8
    \item NVIDIA Corporation GM108M [GeForce 940MX] / NVIDIA GeForce 940MX/PCIe/SSE2 graphical processing unit
    \item running the latest stable version of Ubuntu (20.04.3 LTS), with GNOME version 3.36.8
\end{itemize}

\subsection{A note on the measurement tools}
As mentioned before, both sampling the RAPL interface direcly, as well as CarbonTracker \citep{anthony2020carbontracker} will be used to measure the power/energy consumption. The resulting values should be interpreted differently.

\paragraph{The RAPL interface} measures by directly sampling the power consumption of the CPU every other time interval. This measurement is code-independent. Whether or not your code is running, this method will measure the power consumption. This method gives a good overview of the entire power consumption of some task, including the non-code related parts, e.g. looking up documentation and articles, debugging, running code that finishes with an error, etc. This measurement gives the most accurate measurement of a certain task, but will include variability depending on the efficiency, experience and knowledge of the data scientist performing the task. Note that only the power consumption of the CPU is considered. This may deviate from the true power consumption, especially when running code that relies heavily on RAM. This drawback will be discussed in Section \ref{sec:assumptions_revisited}, when revisiting the assumptions made earlier (see Section \ref{sec:assumptions}).

\paragraph{CarbonTracker} measures the energy consumption of a single piece of code. This measurement gives a good overview of how much energy was consumed by running the code, and nothing but the code. If a script finishes on an error, no measurement will be written out. This measurement has a higher repeatability, but lacks information on the overhead of the tasks: the non-code related parts and running bugged code that finishes with an error.

\paragraph{RAPL vs CarbonTracker} Note that the samples for the Carbontracker measurements is not the same kind of samples as those fore the RAPL measurements. Carbontracker reports aggregate values at the end of every successful exit of a piece of code, independent of how long it took, while RAPL measurements produce a real-time measurement every 10 seconds, independent of what the computer is doing. The variance of the Carbontracker measurements is thus a variance between the total power consumptions of different successfully run code snippets. The variance of the RAPL measurements, on the other hand, is the variance on the real-time power consumption. This difference in sample types should be kept in mind when interpreting the results.

\subsection{Notes on the measurements}
\label{sec:measurements}
The considered measurements on common data science and common computer usage inevitably have a low repeatability due to the specific hardware used, the specific data science project and methods considered, as well as the workflow and work efficiency of the authors. For completeness, the considered measurements are described in as much detail as possible below.

Other reported values for everyday tasks and advanced data science are averages, educated guesses, or can only be reported under certain assumptions. These tasks will also be described in detail below.

\subsubsection{Measurements of regular computer usage}
\label{sec:regular_pc_usage}

\paragraph{Baseline (idle)} This ``task is done by leaving a computer running with the screen on, and no open programs. The value of this task will vary depending on the efficiency of your hardware, background tasks and screen brightness.

\paragraph{Watching a movie} This task was completed by streaming \textit{Avengers: Age of Ultron} via Disney+ on Firefox 96.0 for Ubuntu. No other tabs or programs were running. The carbon emission of the datacenter providing the movie is not taken into account due to lack of available data.

\paragraph{Normal usage} Normal usage entails browsing the internet, reading a pdf and opening/closing various programs: Inkscape, Firefox, text editor and file browser. Tasks that involve video rendering (such as gaming, watching Youtube or viewing local video files) were purposely avoided, as these are comparable to the task \textit{Watching a movie}.

\paragraph{Working in Excel} This task entailed making plots and performing various column transformations on a $4,383\times 20$ dataset containing only continuous numerical values. These column transformations were aimed to reflect common data science operations such as: deleting columns, scaling columns, and power- and log-transformations. Plots were limited to histograms. These Excel operations were performed in LibreOffice Calc version \texttt{1:6.4.7-0ubuntu0.20.04.2}.

\paragraph{Data science project} This task entailed doing a common data science project, namely a credit scoring classification problem: identifying defaulters on a loan. This project is split into two main parts:
\begin{enumerate}
    \item Exploration \& preprocessing
    \item Gridsearch \& fit
\end{enumerate} 
Exploration \& preprocessing entailed constructing plots of features and processing features according to their meaning, type and distribution. This is done by means of column transformations on a $20,000\times35$ dataset with mixed continuous and categorical features. All data exploration and processing was done in Python3.8 using matplotlib, pandas and scikit-learn. Column transformations included: encoding (one-hot encoding and WOE-encoding), scaling to normal distributions with \texttt{QuantileTransformer()}, filling missing values with means or modes and over- and undersampling with ADASYN and \texttt{RandomUnderSampler()}.

\begin{table*}
\centering
\begin{tabular}{lll}
\toprule
Model               & Gridsearch \#1                                                                                                                                                                                          & Gridsearch \#2                                                                                                                                                                 \\ \midrule
Random Forest       & \begin{tabular}[c]{@{}l@{}}n\_estimators: {[}100, 500, 1000, 2000{]},\\ max\_depth: {[}2, 10, 50, 100{]},\\ min\_samples\_split: {[}2, 5, 10, 50{]}\end{tabular} & \begin{tabular}[c]{@{}l@{}}n\_estimators: {[}2000, 3000{]},\\ max\_depth: {[}30, 50, 70{]},\\ min\_samples\_split: {[}2{]}\end{tabular} \\ \midrule
KNN                 & \begin{tabular}[c]{@{}l@{}}n\_neighbors: {[}20, 100, 500, 1000{]},\\ p: {[}1, 2{]}\end{tabular}                                                                                                     & \begin{tabular}[c]{@{}l@{}}n\_neighbors: {[}10, 20, 50{]},\\ p: {[}1{]}\end{tabular}                                                                                       \\ \midrule
Logistic Regression & \begin{tabular}[c]{@{}l@{}}penalty: {[}l2, l1, elasticnet{]},\\ C: {[}0.1, 1, 10, 100, 1000{]},\\ max\_iter: {[}100, 1000, 2000{]}\end{tabular}                                             & \begin{tabular}[c]{@{}l@{}}penalty: {[}l2{]},\\ C: {[}80, 100, 200, 300, 400{]},\\ max\_iter: {[}1500, 2000, 3000{]}\end{tabular}                                      \\ \bottomrule
\end{tabular}
\caption{Hyperparameter grids for the two gridsearches performed during the task \textit{Gridsearch \& fit}.}
\label{tab:hyperparam}
\end{table*}

Gridsearch \& fit entailed performing two hyperparameter gridsearches for the machine learning models Random Forest, KNN and Logistic Regression on the dataset as described in the previous paragraph. Gridsearches were performed by starting with a coarse Gridsearch and refining the grid intervals once with a second Gridsearch. The considered hyperparameter grids are shown in Table~\ref{tab:hyperparam}. A $5$-fold cross-validation scheme was used.

\subsubsection{Measurements of everyday tasks}
\label{sec:everyday_tasks}
\paragraph{Burning a lightbulb for an hour}  Assuming a $10W$ light bulb.

\paragraph{Streaming Season 1 of Friends} The carbon emission of this task was calculated using the same measurement for power consumption as the task \textit{Watching a movie} and extrapolating this to the duration of the first season of Friends ($2h28'$).

\paragraph{Leaving the office lights on over the weekend} This is calculated by assuming the office is lit by $8$ $10W$ TL lights, buring from friday 5pm until monday 9am.

\paragraph{Heating an office for a day} This is calculated by assuming an office with following properties:
\begin{itemize}
    \item $50m^2$ area
    \item One outside wall
    \item Solid and insulated cavity walls
    \item A ceiling height of $3m$
    \item A minimum roof insulation of $75mm$
    \item Electrically heated at $21^\circ C$ ($70^\circ F$)
\end{itemize}
Data was taken from \cite{heating}.

\paragraph{Commercial airflight for 1 passenger (BRU $\xrightarrow{}$ JFK NY)} This carbon emission was calculated assuming a flight from Brussels (BRU) to New York (JFK NY), where all seats are economy seats and occupied. Data taken from \cite{flight}.

\subsubsection{Measurements of advanced data science}
\label{sec:advanced-ds}

\paragraph{Training a CNN on images} This measurement was taken from \cite{anthony2020carbontracker}, where more details can be found. This advanced data science task was performed by training the CNN U-NET for 100 epochs on the LIDC medical image dataset. This required $1.25\pm 0.25\ kWh$, as is visible in Figure 1 of \cite{anthony2020carbontracker}. Note that:
\begin{itemize}
    \item This value was measured using different hardware
    \item The dataset was already processed; processing does not contribute to the reported value
    \item Only one training session is reported, which does not reflect a realistic case of trial and error, with multiple training sessions.
    \item While computationally challenging, it is feasible to replicate this task on a home computer. This measurements lives in the grey zone between common and advanced data science.
\end{itemize}

\paragraph{Training GPT-3 in the EU} This measurement is an estimate taken from \cite{anthony2020carbontracker}, assuming Microsoft's average datacenter PUE (Power Usage Effectiveness) of 2015, re-calculated with updated values for the GHGe associated with energy production (EU28 average from 2020) and GHGe of cars (EU28 average from 2021).
\section{Results}
\begin{figure}
    \centering
    \includegraphics[width=.99\linewidth]{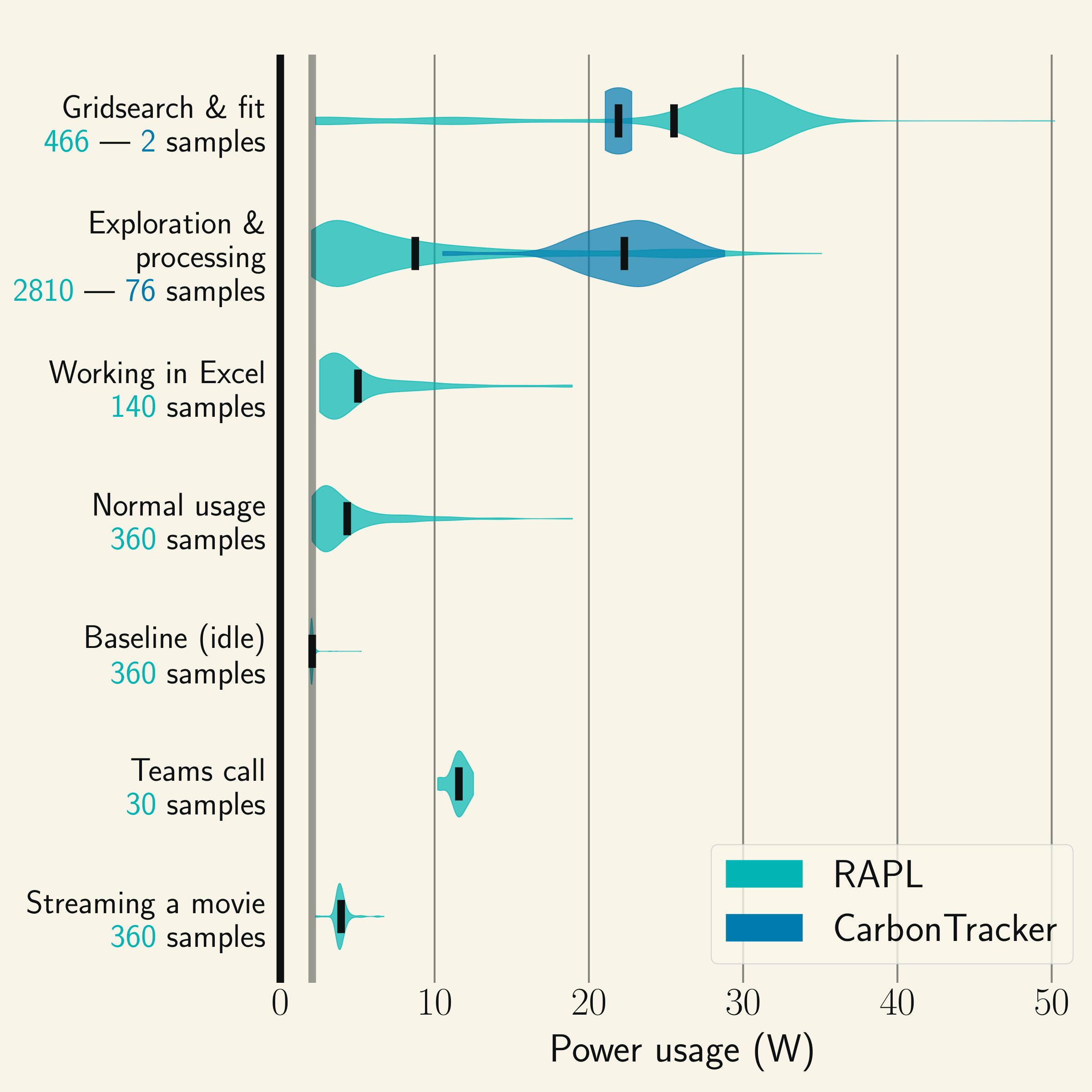}
    \caption{Power consumption distributions of working on common computer tasks and data science, measured by CarbonTracker (dark blue) and sampling the RAPL interface (light teal). RAPL samples are taken every 10 seconds, while CarbonTracker samples are taken every successful run of a piece of code.}
    \label{fig:power}
\end{figure}

Figure \ref{fig:power} shows the power consumption distribution per task.

\begin{figure}
    \centering
    \includegraphics[width=.99\linewidth]{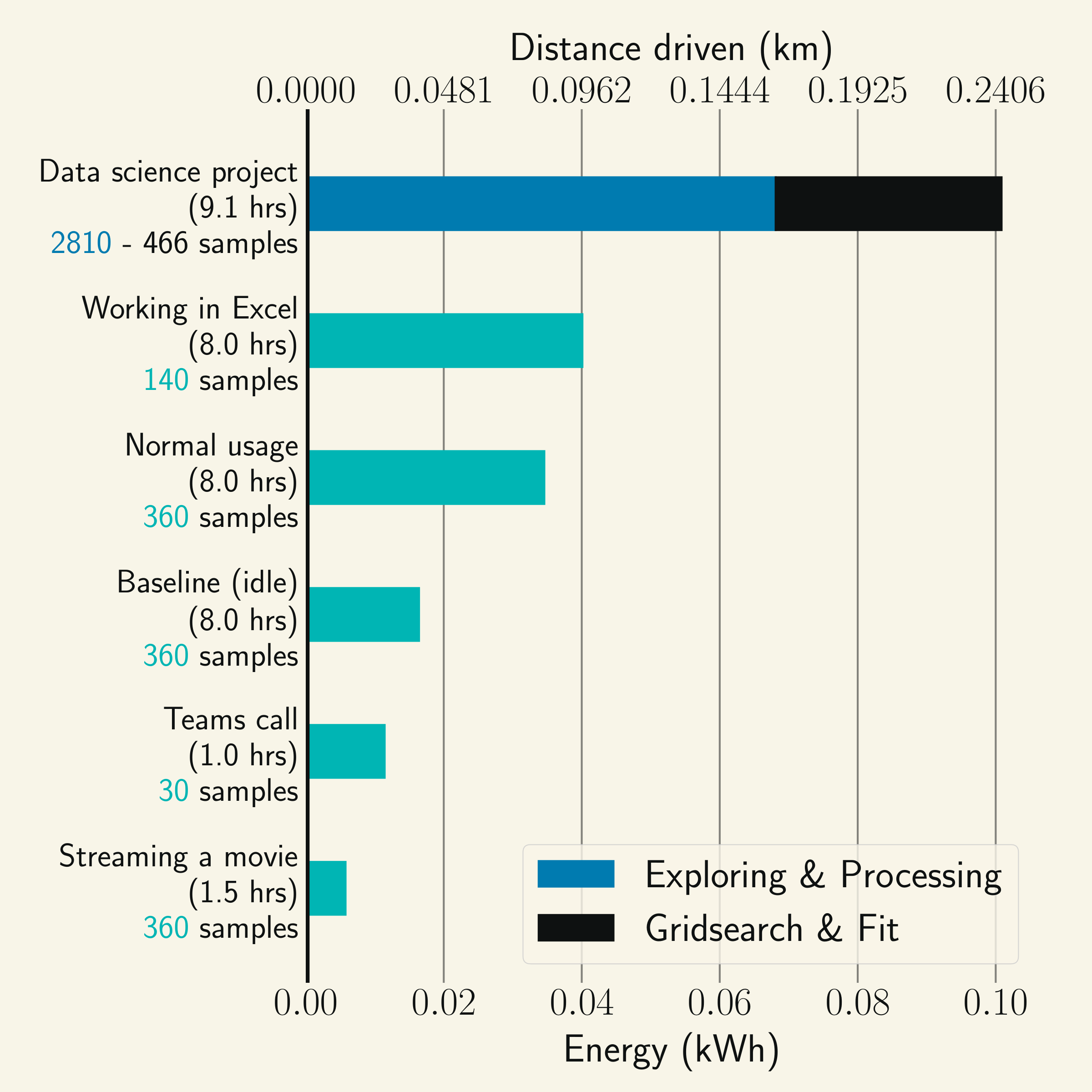}
    \caption{Energy consumption and carbon emission (in km driven by car equivalent) of working on common computer tasks and data science related activities. Only RAPL measurements are shown.}
    \label{fig:energy}
\end{figure}

Figure \ref{fig:energy} shows the energy consumption of working on a certain task. For tasks with an unspecified end time, a duration of 8 hours is used to reflect a day worth of work.

\begin{figure}
    \centering
    \includegraphics[width=.99\linewidth]{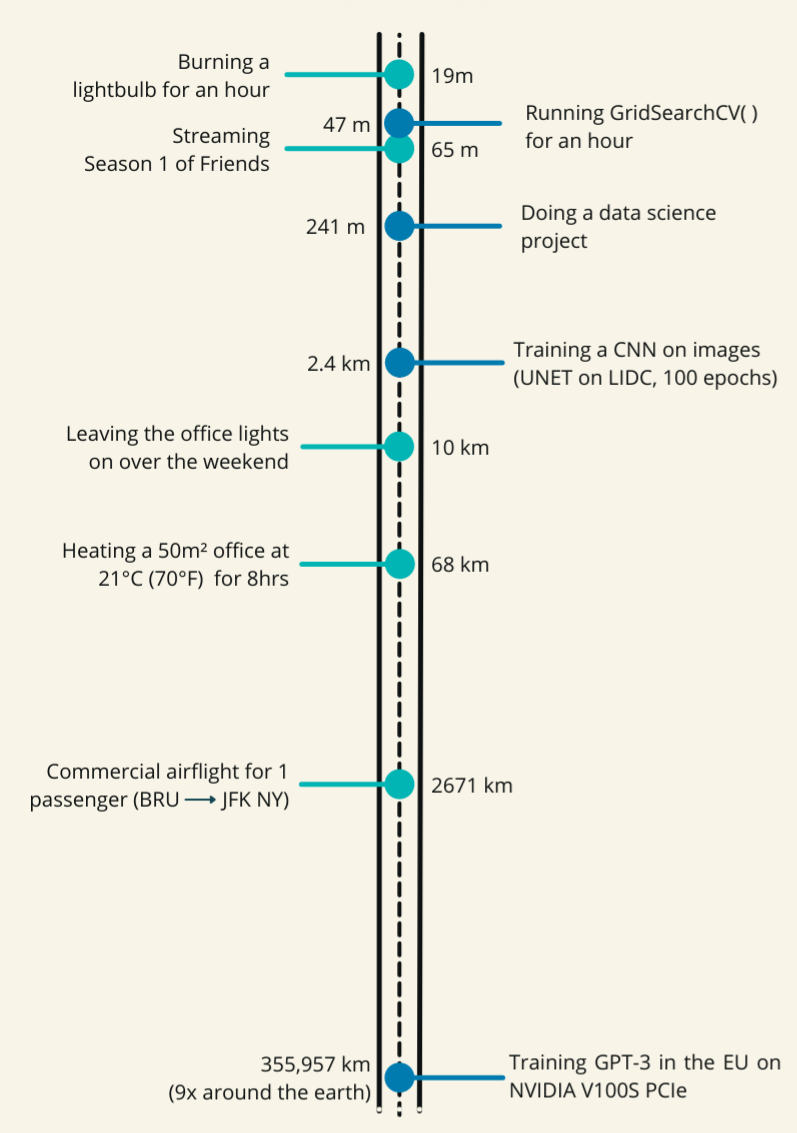}
    \caption{Carbon emission of common non-computer related tasks and data science (common and large-scale) on a logarithmic scale, expressed in units of \textit{km driven by car}.}
    \label{fig:emission}
\end{figure}

These are compared to large-scale data science and other everyday tasks in terms of carbon emission in Figure~\ref{fig:emission}.

\begin{figure}
    \centering
    \includegraphics[width=.99\linewidth]{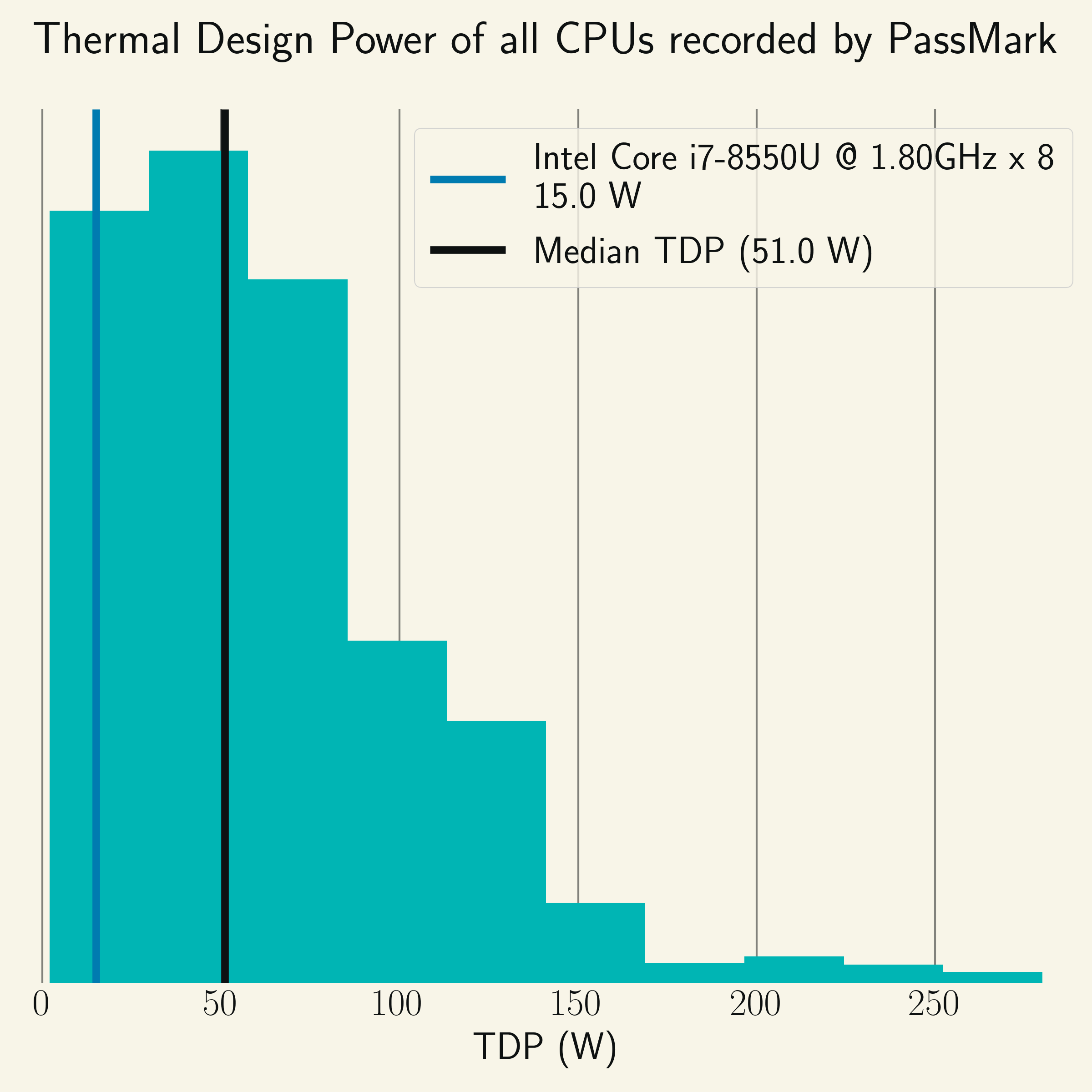}
    \caption{Thermal Design Power of all CPUs as recorded by \protect\cite{passmark}. The TDP of the CPU considered in this work is denoted by a vertical blue line, the median TDP by a black one.}
    \label{fig:TDP}
\end{figure}

The Thermal Design Power (TDP) of the CPU used in this work is compared to other CPUs in Figure~\ref{fig:TDP}.

\section{Discussion}
\label{sec:discussion}

\subsection{Power consumption}
We can see on Figure \ref{fig:power} how CarbonTracker yields similar results on power consumption as directly sampling the RAPL interface for \textit{Gridsearch \& fit}, but there is a notable difference between the two measurements for the task \textit{Exploration \& preprocessing}. This can be explained by the fact that the latter task includes a lot of overhead, such as looking up documentation, scaling columns one by one, programming, debugging and running faulty code. All of the previously mentioned overhead is not captured by CarbonTracker measurements; only code that runs without error is measured by CarbonTracker. CarbonTracker can be interpreted as the hypothetical power consumption of a hyper-efficient data scientist that does not make any mistakes, knows the sklearn library by heart and writes their code instantaneously. \textit{Gridsearch \& fit}, on the other hand, is defined by few to no programming, but simply leaving a script running. This explains the similarity of the measurements of CarbonTracker and directly sampling the RAPL interface.

\subsection{Energy consumption}
Figure \ref{fig:energy} reports the energy consumption of the same tasks as shown in Figure \ref{fig:power}, sorted from low to high. If no duration is reported in the label, a duration of an $8$ hour working day is considered.

It is notable how streaming a $1.5$ hour movie requires less energy than sitting through a $1$ hour Teams call, despite both entailing similar tasks: real-time video rendering. This is possibly due to the fact that online meetings cannot make use of the same video compression methods as streaming a movie, since the video data of the former is produced in real-time. Processing less compressed data would then require more computational power and put the CPU under heavier load, requiring more energy. It must be noted that the carbon emission of the datacenter providing the movie is not taken into consideration. This would increase the associated carbon emission.

It is also notable how the $2$ gridsearches make up about $1/3$ of the total energy consumption of the data science project, despite making up only $14\%$ of the time ($1h17'$ out of $9h04'$). This makes sense, as performing a gridsearch makes a CPU run in parallel under heavy load, yielding a high power throughput. This was already visibli in Figure \ref{fig:power}.

\subsection{Carbon emission}
\label{sec:results-emission}
Figure \ref{fig:emission} shows a log-scale of the carbon emission of various everyday tasks, common data science tasks, and advanced data science tasks. The carbon emission is expressed in units \textit{km driven by car} as described in Section \ref{sec:converting}.
It is visible how all considered common data science tasks emit less $CO^2eq$ than e.g. leaving the office lights on over the weekend. Even an advanced data science task such as training a CNN on images emits less $CO^2eq$, when performed under the same conditions as described by \cite{anthony2020carbontracker}. Note that these conditions are specific and impose great variance. For example, training your model in Estonia will yield about $64$ times more $CO^2eq$ emission than training it in Iceland, and training for twice as many epochs will yield an emission about twice as big. The measurement is supposed to reflect a representative training session.

\subsection{Hardware}
Figure \ref{fig:TDP} shows a distribution of the Thermal Design Power (TDP) of all CPUs recorded by \cite{passmark}. The distribution does not represent the relative frequency of these CPUs; each CPU is a single datapoint, independent of its popularity in modern hardware. It shows how the CPU used in this work is more efficient than most CPUs, yielding a TDP of $15.0W$ compared to the median TDP of $51.0W$. Measurements of the tasks as described in Section \ref{sec:regular_pc_usage} may yield higher values when performed on other hardware.

\subsection{Assumptions revisited}
\label{sec:assumptions_revisited}
Let us revisit the assumptions made in Section \ref{sec:assumptions} and verify if they hold true, or in which way they should be adapted. They are numbered in the same way as in Section \ref{sec:assumptions}
\begin{enumerate}
\item \cite{anthony2020carbontracker} shows that the RAM can easily make up to $50\%$ of the power consumption during the training of a neural network. Measuring only the power consumption of the CPU is not representative, especially during RAM-needy code. It can be estimated that the results associated with the hardware-specific tasks as described in Section \ref{sec:regular_pc_usage} are, at worst, only half of the true power consumption.
\item The CPU used in this work may not be very representative for other hardware architectures. If we assume the median TDP of all CPUs recorded by \cite{passmark} is more representative for other CPUs, then the results associated with the hardware-specific tasks as described in Section \ref{sec:regular_pc_usage} are $3.4\times$ too small. The latter assumption, i.e. assuming the difference between CPU TDP values is representative for the difference in realistic power consumption across CPUs in various hardware structures, is yet another dangerous assumption. TDP values should be interpreted carefully, as they do not reflect realistic power consumption:
\begin{itemize}
\item The TDP values from the manufacturers are often incorrect \citep{TDPinaccurate}.
\item They only apply when the CPU is under full load on all cores, which is rare.
\item Actual power usage can be altered when the computer is running via power plans (e.g. switching to battery power).
\item They don't take into account the electrical usage of other components, such as the screen, hard drive, RAM, etc.
\end{itemize}

In combination with the previous assumption about hardware, it is thus possible that a more realistic power consumption (and associated carbon emission) of the hardware specific tasks in this work can be up to $6.8\times$ larger when including an estimate for the RAM power consumption. The relative difference between regular computer usage and common data science as shown in Equation~\ref{rel_diff} is not influenced by this systematic error.

\item A comparison between the author's efficiency and methods to other data scientists is next to impossible to measure. This assumption retains its status of hypothesis.

\item Looking at the RAPL measurements of power consumption shown in Figure \ref{fig:power}, it is visible how the lower three measurements have very low variance, and the remaining measurements have a well-defined `blob'. The former indicates that the tasks had a low variance in power consumption throughout time and this was successfully captured with the considered sampling rate of $10s$. The latter indicates that the most frequent value for power consumption (the `blob') was successfully captured by the considered sampling rate, but outliers may not have been accurately measured. If the power consumption varied too quickly, a sampling rate of $10s$ was too large to properly capture this behaviour. As the blob 
is well-defined for each task, the general behaviour of the time-dependency of the power consumption is well captured, and this effect is not considered as substantial.

\item The conversion of $gCO^2eq$ to \textit{km driven by car} is based on the EU average from 2019. \cite{carco2} shows a clear declining trend in car emission. The values expressed in the latter unit in this paper will become increasingly outdated every year. Of course, due to the variance in carbon emission for each car, the values also vastly differ depending on which car you consider.

\item As already mentioned at the end of Section~\ref{sec:results-emission}, the conversion from energy to $gCO^2eq$ depends heavily on which country you consider. Outsourcing heavy computational work to countries with a greener energy production can substantially decrease the carbon footprint.
\end{enumerate}

To estimate the additional contribution of common data science to the global GHGe, let's use the results and the revisited assumptions to compare common data science to regular computer usage. When using the RAPL measurements, and scaling everything to the same duration, and under the assumptions that:
\begin{enumerate}
    \item the hardware-specific power consumption in this work can be extrapolated linearly to other hardware configurations, independent of the power load.
    \item the data science project and normal computer usage in this work are representative for other computer users, data scientists and data science in general.
\end{enumerate}
The extra load in global GHGe of common data science when compared to normal computer usage for a single device is equal to
\begin{equation}
\label{rel_diff}
    \begin{aligned}
    \frac{E_{data science Project}}{E_{Normal Usage}} &= \frac{8\ hr}{9.1\ hr} \frac{100.9955\ Wh}{34.72\ Wh} \\
    &= 2.57
    \end{aligned}
\end{equation}

In order to estimate the additional load in GHGe due to common data science on a global scale, rather than for a single device, one would have to link this value to the relative global GHGe of household ICT devices of $~45\%$ \citep{ICTGlobalEmission}. To do so, one would need a value for how many of these devices are used for data science (and how often), which is very hard to obtain.

\section{Conclusion}
The results made clear that common data science (i.e. any form of data science that can be performed on your laptop at home) requires substantially more power than other common computer tasks, yielding an associated carbon footprint that's about $2.57$ times higher than normal computer usage. Keeping in mind that data science jobs have been on a steady popularity increase over the last few years, this implies a higher global GHGe associated with common data science.

When comparing common data science to advanced data science, it is clear how the power consumption increases exponentially along with the complexity of the data science. While advanced data science tasks, such as training GPT-3, require a substantial amount of power, this is not the case for common data science.

The discussion revealed that the measurements for common data science in this work can be up to $6.8$ times larger than reported. Even with this value in mind, common data science  does not appear to contribute substantially to the global carbon emissions compared to other everyday tasks. While the uprise in data science jobs can indeed be associated with an increased carbon footprint, it will prove much more substantial to be mindful about the power consumption and associated carbon emission of some everyday tasks. Efficient heaters and lighting, insulation, green energy production and green travel methods will always beat, by a landslide, preferring e.g. a \texttt{RandomsearchCV()} over a \texttt{GridsearchCV()} when doing common data science.

That does not mean, however, that efforts to reduce the carbon footprint of data science are in vain. Common data science may be easier on the energy bill than bad insulation, but the computational needs of data science in general, including advanced data science, scale along with their complexity. If mindfulness about the carbon footprint of data science becomes ubiquitous, independent of the scale of the data science project, the difference in GHGe will scale accordingly. The bigger the model, the bigger the difference.

\bibliography{bib}
\end{document}